\input harvmac
\Title{ \vbox{\baselineskip12pt
\hbox{hep-th/9912152}
\hbox{HUTP-99/A073}
%\hbox{IASSNS-HEP-99-52}
\hbox{NUB-3207}}}
{{\vbox{
\centerline{RR Flux on Calabi-Yau and Partial}\bigskip
\centerline{ Supersymmetry Breaking}}}}\
\smallskip
\centerline{Tomasz R. Taylor}
\smallskip
\centerline{\it Department of Physics, Northeastern University}
\centerline{\it Boston, MA 02115, USA}
\bigskip
\centerline{Cumrun Vafa}
\smallskip
\centerline{\it Jefferson Laboratory of Physics}
\centerline{\it Harvard University, Cambridge, MA 02138, USA}\bigskip

\bigskip

\def\tilde{\widetilde}

\def\F{{\cal F}}
\def\N{{\cal N}}
\def\Im{{\rm Im}}
\def\Re{{\rm Re}}
\def\pl{Phys. Lett. B~}
\def\np{Nucl. Phys. B~}
\def\pr{Phys. Rev. D~}
\vskip .3in
We show how turning on Flux for RR (and NS-NS)
field strengths on non-compact Calabi-Yau 3-folds can serve as a way to
partially break
supersymmetry from $N=2$ to $N=1$ by mass deformation.  The freezing of
the moduli
of Calabi-Yau in the presence of the flux is the familiar phenomenon of
freezing of fields in supersymmetric theories upon mass deformations.

\def\underarrow#1{\vbox{\ialign{##\crcr$\hfil\displaystyle
 {#1}\hfil$\crcr\noalign{\kern1pt\nointerlineskip}$\longrightarrow$\crcr}}}
% use of underarrow
%A~~~\underarrow{a}~~~B
\Date{December 1999}

\newsec{Introduction}

Type II strings compactified on Calabi-Yau threefolds
give rise to $N=2$ theories in 4 dimensions.  The
geometry of Calabi-Yau threefold and its moduli space
provides a deep insight into the dynamics of $N=2$
gauge theories.  It is thus natural to ask if the
simple operation of breaking supersymmetry from $N=2$ to $N=1$
(say by addition of mass terms)
has a Calabi-Yau counterpart.  If so, this may provide
insight into the dynamics of $N=1$ gauge theories.

A particular approach to breaking supersymmetry
in the context of Type II compactification on Calabi-Yau
threefolds was taken in \ref\post{J. Polchinski and A. Strominger,
\pl 388 (1996) 736.}\ where
 Ramond-Ramond fields strengths were turned on.
It was shown, however, that one either preserves
all $N=2$ supersymmetries (and freeze the moduli
of Calabi-Yau to make it correspond to singular limits
such as the conifold) or one breaks the supersymmetry
completely.  Furthermore it was argued in \ref\jm{J. Michelson,
\np 495 (1997) 127.}\
that this is a general result.

On the other hand it was found in 
\ref\apt{I. Antoniadis, H. Partouche and T.R. Taylor,
\pl 372 (1996) 83; I. Antoniadis 
and T.R. Taylor, Fortsch.\ Phys. 44 (1996) 487;
H. Partouche and B. Pioline, Nucl. Phys. Proc Suppl. 56B (1997) 322.}\
that in the context of $N=2$  quantum field theories
it is possible to add $N=2$ FI terms, and break the supersymmetry
to $N=1$. These constructions were generalized to the local
case in \ref\fgp{S. Ferrara, L. Girardello and M. Porrati, \pl 376 (1996) 275;
M. Porrati, Nucl. Phys. Proc. Suppl. 55B (1997) 240.}\ (see also 
\ref\bg{J. Bagger and A. Galperin, Phys. Rev. D~55 (1997) 1091;
M. Ro\v{c}ek and A. A. Tseytlin, Phys. Rev. D~59 (1999) 106001.}).
There seemed, therefore, to exist a conflict between
the results coming from considerations of type II compactifications
on Calabi-Yau threefolds which suggested finding $N=1$ supersymmetric
theories by turning on fluxes is not possible, whereas field
theory arguments suggested that some such deformations should
be possible.

We will see in this paper that indeed we can obtain
partial supersymmetry breaking by considering non-compact
Calabi-Yau manifolds with fluxes turned on.  The way
this avoids the no-go theorem in \jm\ is by taking
a certain decompactification limit, which renders some
fields non-dynamical.  In other words, it would have corresponded
to a theory with no supersymmetric vacua in the compact situation,
and where it not for making some fields non-dynamical, we could
not have obtained partial supersymmetry breaking.  However, as far
as geometric engineering of $N=1$ theories are concerned
the non-compactness of
the Calabi-Yau is a perfectly acceptable condition, and this
is already the case for geometric engineering of $N=2$ theories.

The organization of this paper is as follows:
In section 2 we show how supersymmetry can be partially
broken by considering a simple generalization of models
of \apt\ where we include two $N=2$ vector multiplets.
In section 3 we consider type II compactifications on 
Calabi-Yau threefolds and show why turning on RR fluxes (and
in addition
NS flux $H$ for type IIB) is equivalent to turning on FI terms in the
$N=2$ supersymmetric theory.  We also review the no-go theorem
of \jm\ and show how it may be
avoided in certain non-compact  limits.  In section 4 we briefly
review aspects of $N=2$ geometric engineering and
show how RR flux can give mass to the adjoint field breaking
the theory to $N=1$ and freezing some of the Calabi-Yau
moduli.

Main results of this paper have also been obtained by Peter Mayr
\ref\maryp{ P. Mayr, to appear.}.  Related ideas have also been considered
in \ref\kouki{E. Kiritsis and C. Kounnas, \np 503 (1997) 117.}.

\newsec{Partial Supersymmetry Breaking and Mass Generation}
In this section, we present a simple generalization of the model
discussed in \apt\
 which exhibits partial supersymmetry
breaking with mass generation for $N=1$ multiplets. 
It involves two $N=2$ vector multiplets, $S$ and $A$, with the prepotential
\eqn\prep{ \F(S,A)= {i\gamma\over 2} S^2 + {1\over 2}SA^2\,  , }
where $\gamma$ is a real constant, and with the superpotential
(which in general can be taken to be a linear combination of ``periods'')
\eqn\superpot{ W~=~eS+m\F_S~=~(e+i m\gamma)S+{m\over 2}A^2\, , }
where $e=e_1+ie_2$ and $m=m_1+im_2$ are {\it complex\/} constants
and $\F_S=\partial \F /\partial S$.
The corresponding Lagrangian
is $N=2$ supersymmetric. The constants $e$ and $m$ correspond to
$N=2$ electric and magnetic Fayet-Iliopoulos terms, respectively. 
In the manifestly $N=2$ supersymmetric notation of Ref.\apt :
\eqn\em{\Re\vec{E}=(e_1~ e_2~ 0)\ ,\qquad\vec{M}=(m_1~ m_2~ 0)\, .}

The superpotential of Eq.\superpot\ gives rise to the following potential
for the scalars $S=\alpha+i\sigma$ and $A=b+ia\,$:
\eqn\vapt{V={|\, ea-\gamma m b\,|^2\over \gamma (\gamma \sigma-a^2)}.}
In the above equation, we neglected an irrelevant, additive constant term.
The potential has a zero-value minimum at $a=b=0$. 
For generic values of $e$ and $m$, both $N=2$ supersymmetries are
broken spontaneously. There are, however, two special 
configurations of these parameters:
\eqn\emc{ e\pm im\gamma=0\, ,}
for which supersymmetry is broken partially to $N=1$. In this case,
both scalars $a$ and $b$, as well as the fermionic component of the
$N=1$ chiral multiplet $A$ acquire an equal mass of $|m|/\sigma$.
The simplest way to prove formally that such a partial breaking does indeed
occur is to follow the method of \apt\ and examine the supersymmetry
variations of fermions. In this way, one can identify the $N=2\to N=1$ 
goldstino as one of the two fermionic components (gauginos) of the
$S$ multiplet. In fact, the full $N=2$ vector multiplet $S$ 
and the $N=1$ vector component of $A$ remain massless
while the $N=1$ chiral multiplet $A$ acquires a mass.

The above model can be generalized to more complicated prepotentials, 
of the form
\eqn\fgen{\F(S,A)=f(S)+{1\over 2}SA^2.}
As in the previous case, the potential has a minimum at $A=0$.
However, there is also another minimization, with respect to $S$,
which yields two solutions
\eqn\aptvac{e+m\F_{SS}=0 \qquad{\rm or}\qquad e+m\bar{\F}_{\bar{S}\bar{S}}=0\, ,}
similar to \emc .
It is easy to see that the above equation is exactly the condition for partial
supersymmetry breaking.
Hence we conclude that an $N=1$ supersymmetric vacuum
exists also in the general case. In particular, the mass $|m|/\sigma$
is generated again for the $N=1$ chiral multiplet $A$. 

So far we have been discussing the case of global
$N=2$ supersymmetry.  In the context of string theory
we of course have local $N=2$ supersymmetry.  In such
a case to obtain $N=2$ global limit we have to take
some particular limit, where gravity decouples,
say by taking in the type II context weak limit
of string coupling constant, and perhaps some other
limits for other fields.
In this context we can break $N=2$ to $N=1$ in an
even simpler way.
  Set $\gamma =0$, so that the prepotential is just
$${\cal F}={1\over 2}SA^2$$
This would have given a singular kinetic term for $S$ in
the global case, but it is perfectly fine in the local case.
We can think of $S$ for example
as the ``heterotic string coupling
constant''.  We now turn on
 FI term  $\alpha {\cal F}_S+\beta {\cal F}_A$. 
 We take the limit where the vev of $S$ becomes large
(i.e. weak coupling heterotic string limit).  In this limit $S$ becomes
non-dynamical.  And the superpotential term $W={1\over 2}m(A')^2$ 
(where $A'$ is related to $A$ by a shift) simply
gives mass to the scalar $A'$, breaking $N=2$ to $N=1$. It
is this realization of partial supersymmetry breaking
  that we will find applicable
in the Calabi-Yau context later in this paper.

\newsec{Type IIB on Calabi-Yau 3-fold with  $H$-flux}
Consider compactification of type IIB on a Calabi-Yau
threefold. We would like to consider turning on flux
for NS and R threeform field strengths $H^{NS}$ and $H^R$.
This is a case already considered in \jm\
following the work of \post\
and more recently from the viewpoint of F-theory in\nref\gvw{S. Gukov, C. Vafa
and E. Witten, hep-th/9906070.}\nref\setet{K. Dasgupta,
G. Rajesh and S. Sethi, JHEP 9908 (1999) 023.} \refs{\gvw,\setet}.
The theory has $h^{2,1}$ vector multiplets
and $h^{1,1}+1$ hypermultiplets in addition to the $N=2$ gravitational
multiplet, where $h^{p,q}$ denotes Hodge numbers of Calabi-Yau.
The relevant modification to the effective action
due to turning on $H$-flux is in interactions with the vector
multiplets.  Let $\Omega$ denote the holomorphic threeform
on the Calabi-Yau.  
We write the effective Lagrangian we obtain in 4 dimensions
in an $N=1$ supersymmetric framework.  The net effect
of
turning on $H$-flux is to add a superpotential
of the form
\eqn\supp{W=\int \Omega \wedge (\tau H^{NS}+ H^{R})}
in the 4-dimensional effective theory,
where $\tau $ denotes the complexified coupling
constant of type IIB strings.  Note that $H^{NS}$
and $H^R$ are dual to some integral 3-cycles
$C_{NS}$ and $C_R$ and the above formula can also be
written as
$$W=\int_{\tau C_{NS}+C_R}\Omega$$
To see how \supp\ arises note
that if we consider a five brane (NS or R) wrapped around
a 3-cycle $C$ in the Calabi-Yau, it corresponds
to a domain wall in 3+1 dimensional theory, whose
BPS bound for tension should be given by
$\Delta W$ across the domain wall. On the other hand
the tension of the 5-brane should be $\int_C\Omega$
(times $\tau$ in the case of NS 5-brane).
Since the 5-brane wrapped around $C$ changes the $H$ flux
across the domain wall
by a 3-form dual to the $C$ cycle we see that this
gives
the expected change $\Delta W$.  This argument was discussed
in \gvw\ in the context of F-theory on 4-folds, and type IIB
on Calabi-Yau 3-folds is a special case of it.

We can also write \supp\ explicitly
 if we choose a basis for $H_3(M,{\bf Z})$,
given by $(A^{\Lambda},B_{\Sigma})$, $\Lambda,\Sigma=0,...,h^{2,1}$,
with $A^{\Lambda} \cap A^{\Sigma}=
B_{\Lambda}\cap B_{\Sigma}=0$ and $A^{\Lambda} \cap B_{\Sigma}=
\delta^{\Lambda}_{\Sigma}$.
Sometimes we refer to $A^{\Lambda}$ as the electric cycles and $B_{\Sigma}$ as
the magnetic cycles. This clearly is a basis dependent definition.
Let
$$X^{\Lambda}=\int_{A^{\Lambda}}\Omega \qquad \F_{\Sigma}=
\int_{B_{\Sigma}}\Omega$$
Moreover denote the dual 3-cycle to the H-fluxes
by 
$$\tau C^{NS}+C^R=e_{\Lambda} A^{\Lambda}+m^{\Lambda} B_{\Lambda}$$
where
\eqn\cs{C^{NS}=e^1_{\Lambda}A^{\Lambda}+m^{1\Lambda}B_{\Lambda}
 \qquad\qquad C^{R}=e^2_{\Lambda}A^{\Lambda}+m^{2\Lambda}B_{\Lambda}\ ,}
and the complex vectors $e$ and $m$ are defined as:
\eqn\eemm{e_{\Lambda}=e^1_{\Lambda}\tau+e^2_{\Lambda}\qquad\qquad
{}~~~m^{\Lambda}=m^{1\Lambda}\tau+m^{2\Lambda}\, .}
The superpotential \supp\ can be written explicitly as
\eqn\suexp{W=\int_{C^R}\Omega +\tau \int_{C^{NS}}\Omega =e_{\Lambda}X^{\Lambda}
+m^{\Lambda}\F_{\Lambda}}
As is well known there is a prepotential $\F(X)$,
a homogeneous function of weight $2$ in $X$  in terms of which
$$\F_{\Lambda}=\partial_{\Lambda}\F\, .$$
Thus the FI terms are realized by $H$ fluxes in type IIB
string compactification on Calabi-Yau threefolds.

\subsec{Type IIA version}
The same analysis can be done in the type IIA language
(for the case of type IIA on Calabi-Yau 4 folds see
\ref\guko{S. Gukov, hep-th/9911011.}).  In
fact mirror symmetry already tells us what the story will
be in the type IIA case.  The story is much simpler
in the context of just turning on the $H^R$ flux.  In this case
the mirror corresponds to turning on $F^2,F^4 $ and $F^6$ fluxes
which are dual to $4,2$ and $0$ cycles on Calabi-Yau 3-fold.
The analog of \supp\ is now
$$W=N_0+\int_{C_2}k+\int_{C_4}k^2$$
where $k$ denotes the 
K\"ahler class on the Calabi-Yau threefold
and $N_0$ denotes the quantum of $F^6$ flux.
The above formula receives world sheet instanton correction
as is well known, and in fact by mirror symmetry one can
recover the instanton corrected superpotential $W$
on the Calabi-Yau. 

\subsec{Scalar Potential}

Our next step is to obtain the scalar potential corresponding
to the superpotential \suexp . We would like to maintain a
manifest $N=2$ supersymmetry, however this is not possible in the
locally supersymmetric case because the superpotential is 
a genuinely $N=1$ quantity. Instead of turning
to the fully-fledged $N=2$ supersymmetric formalism 
\ref\af{L. Andrianopoli, 
M. Bertolini, A. Ceresole,
R. D'Auria, S. Ferrara, P. Fr\'e and T. Magri, J. Geom. Phys. 23 
(1997) 111.}\ (like in Ref.\jm ), we
can try to obtain the potential by compactifying the 
10-dimensional action. Alas, this is not so simple in view of the
absence of a fully covariant, off-shell formulation of  type IIB supergravity. 
The best we can do is to start from the ``non-self-dual'' (NSD) action 
\ref\bbo{E. Bergshoeff, H.J. Boonstra and T. Ortin, \pr 53 (1996) 7206.}\ 
employing a 4-form field strength which is 
not self-dual. The equations of motion of IIB supergravity follow from the
NSD action after imposing the self duality constraint at the level
of field equations.
Using the NSD action to determine the scalar potential is somewhat 
questionable, nevertheless it is interesting to compare the result with the
superpotential \suexp. In fact, this method will provide
an independent derivation of \suexp. 

In order to parameterize the $H$-fluxes, we will use
the $H^{3}(M,{\bf Z})$ basis $(\alpha_{\Lambda},\beta^{\Sigma})$, 
dual to the $(A^{\Lambda}, B_{\Sigma})$ basis of $H_{3}(M,{\bf Z})$, with
$\int\alpha_{\Lambda}\wedge\beta^{\Sigma}=\delta^{\Sigma}_{\Lambda},~
\int\alpha_{\Lambda}\wedge\alpha_{\Sigma}=
\int\beta^{\Lambda}\wedge\beta^{\Sigma}=0$.
The fluxes can be written as
\eqn\hflux{ H^{NS}=e^1_{\Lambda}\beta^{\Lambda}+
m^{1\Lambda}\alpha_{\Lambda}~ ,\qquad
H^{R}=e^2_{\Lambda}\beta^{\Lambda}+m^{2\Lambda}\alpha_{\Lambda}\, .}
In the presence of the fluxes, the 10-dimensional kinetic
terms give rise to the potential:
\eqn\vflux{ V=(2\Im\tau)^{-1}\int
(\tau H^{NS}+H^R)\wedge *(\bar{\tau} H^{NS}+H^R).}
The integration over the Calabi-Yau manifold can be performed
by using standard
properties of $(\alpha_{\Lambda},\beta^{\Lambda})$ basis
(see e.g.\nref\s{H. Suzuki, Mod. Phys. Lett. A 11 (1996) 623.}
\nref\caf{A. Ceresole, R. D'Auria and S. Ferrara, Nucl. Phys. Proc. Suppl. 46
(1996) 67.}\refs{\s,\caf}), with the result 
\eqn\vh{ V=-(2\Im\tau)^{-1}[m(\Im\N)\bar{m}+(e+m\Re\N)(\Im\N)^{-1}(\bar{e}
+\bar{m}\Re\N)]\, ,}
where $\N$ is the period matrix \caf\ while $e$ and $m$ are the complex vectors
defined in \eemm. The potential can be rewritten as
\eqn\vlo{ V=-(2\Im\tau)^{-1}[(e+m\bar{\N})(\Im\N)^{-1}(\bar{e}+\bar{m}
{\N})]
+m\times e\, ,}
where the constant term 
\eqn\mte{m\times e\equiv m^{1\Lambda}e^2_{\Lambda}-
m^{2\Lambda}e^1_{\Lambda}\, .}
As we will discuss in the next subsection $m$ and $e$ should
be chosen so that $m\times e$ is zero
(for cancellation of 3-brane tadpoles), which we will assume is the case.

In order to relate the above potential with the superpotential \suexp,
we first use the identity \caf :
\eqn\nrel{
e^{-K(z,\bar{z})}(\Im\N)^{-1\Lambda\Sigma}=-2\bar{X}^{\Lambda}X^{\Sigma}-
2D_i{X}^{\Lambda}\,
G^{i\bar{\jmath}}\,D_{\bar{\jmath}}\bar{X}^{\Sigma}\, ,}
where $K$ is the K\"ahler potential of the $N=2$ vector multiplet moduli $z_i$,
$i=1,\dots,h^{2,1}$, and $G^{i\bar{\jmath}}$ is the inverse metric
on the vector moduli space. 
The above expression contains the K\"ahler covariant derivatives:
\eqn\kder{
D_i{X}^{\Lambda}=(\partial_i+K_i){X}^{\Lambda}.}
By using the relations
\eqn\xrel{
\N_{\Lambda\Sigma}X^{\Sigma}=\F_{\Lambda}~,\qquad
\bar{\N}_{\Lambda\Sigma}D_iX^{\Sigma}=D_i\F_{\Lambda}}
we can rewrite the potential as
\eqn\potential{
V=e^{[K(z,\bar{z})+\widetilde{K}(\tau,\bar{\tau})]}\big[G^{i\bar{\jmath}}D_i{W}
D_{\bar{\jmath}}\overline{W}+\widetilde{G}^{\tau\bar{\tau}}D_{\tau}{W}
D_{\bar{\tau}}
\overline{W}\big] ,}
where $W$ is the superpotential \suexp.
The dilaton K\"ahler potential is $\widetilde{K}(\tau,\bar{\tau})=
\qquad\qquad -\ln[(\tau-\bar{\tau})/2i]$ and accordingly,
\eqn\tder{
D_{\tau}{W}=(\partial_{\tau}+\widetilde{K}_\tau)W~,\qquad 
\widetilde{G}^{\tau\bar{\tau}}=\widetilde{K}_{\tau\bar{\tau}}^{-1}=
-(\tau-\bar{\tau})^2.}

Eq.\potential\ is very similar to the standard $N=1$ supergravity
formula for the potential. However,
it is not exactly the same: for instance, the ${-}3|W|^2$ term is missing.
This apparent discrepancy has a simple explanation. 
Although the superpotential does not depend on hypermultiplets,
except on the dilaton $\tau$, the potential receives contributions
from the K\"ahler covariant derivatives\foot{For simplicity, we assume here
that the quaternionic hypermultiplet manifold is K\"ahler.}
with respect to chiral components of all hypermultiplets,
including the Calabi-Yau volume etc. All these contributions are proportional
to $|W|^2$ and must cancel the ${-}3|W|^2$ term. The coefficient
$-3$ is related to the fact that the 4d coupling is rescaled
by the volume of the internal Calabi-Yau threefold.

Now we consider the rigid supersymmetry limit of \potential\ and \suexp.
The Weyl rescaling of the metric that
restores the Planck mass $M_{Pl}$ in the action introduces the factors $M_{Pl}^2$
in front of the Ricci scalar $R$ and scalar kinetic terms; 
the potential acquires a factor of $M_{Pl}^4$. 
Gravity decouples in the $M_{Pl}\to\infty$ limit and the only scalars surviving
as dynamical fields are those with the K\"ahler metric
$\sim M_{Pl}^{-2}$; all other scalars ``freeze'' and can be treated as
constant parameters. In order to recover
the globally supersymmetric models of the type discussed in Section 2 and 
in Ref.\apt\ from type IIB theory with $H$-fluxes, we  
scale the holomorphic sections as $(X^0,\F_0)\sim 1$ and 
$(X^\Lambda,\F_{\Lambda})\sim M_{Pl}^{-1}$, $\Lambda>0$. 
The K\"ahler potential $K(z,\bar{z})$ 
scales then as $M_{Pl}^{-2}$, so the vector moduli survive
as dynamical fields in the $M_{Pl}\to\infty$ limit while $\tau$ and
other hypermultiplets decouple and can be treated as (complex) parameters.
Furthermore, we set $e_0=m^0=0$ and adjust the remaining Fayet-Iliopoulos
parameters so that the superpotential \suexp\ scales as $ M_{Pl}^{-3}$.
In this way \af , we obtain from \potential\ a finite potential corresponding
to the rigid superpotential $W=e_iz^i+m^i\F_i$ \apt. The procedure of taking 
the rigid limit can be further refined to treat some vector 
moduli in  a special way (such as the $S$ field discussed
in the previous section), in order to
freeze them in a way similar to $\tau$. This will be useful in the
context of geometric engineering of $N=1$
theories, to be discussed in the next
section.

\subsec{Supersymmetric Vacua}
Now we analyze supersymmetric solutions with superpotential
given by \suexp . The condition for getting a supersymmetric
solution with $R^4$ background in this context has been studied
by \jm\ with the conclusion that either there are no supersymmetric
vacua or that the $N=2$ is preserved at the vacua.  In particular
no $N=1$ supersymmetric vacua were found in this way.  Let us review
these results in the $N=1$ setup that we are considering.  The condition
for finding supersymmetric vacua in $R^4$ background is that 
$$W=dW=0$$
where $dW$ denotes the derivative of $W$ with respect
to all chiral fields.  In the context of compact Calabi-Yau,
considered in \jm , turning on both $H^{NS}$ and $H^R$
can preserve supersymmetry only if 
\eqn\conddd{\int H^{NS}\wedge H^R \sim m\times e=0.}
Otherwise these fluxes induce anti-3-brane charge in the
uncompactified spacetime (proportional to $\int H^{NS}\wedge H^R$)
and to cancel it we will necessarily break supersymmetry.  If $H^{NS}$
and $H^R$ satisfy \conddd\ then we can choose both of them be dual
to some A-cycles (i.e. m=0).  If we denote the dual three cycles
by $N_1A_1$ and $N_2 A_2$ with periods $\int_{A_i}
\Omega=X_i$ the superpotential will take the form,
$$W=N_1 \tau X_1+N_2 X_2.$$
The condition that $W=dW=0$ in terms of physical fields $z_i=X_i/X_0$,
is equivalent to $dW=0$ in terms of the $X_i$ variables. Since
$X_1$ and $X_2$ are independent fields, we see that condition
$dW/dX_1=dW/dX_2=0$  has no solutions, and so supersymmetry is completely
broken. 

So we see that if we consider smooth Calabi-Yau manifolds
there are no supersymmetric solutions.  However, near singular
Calabi-Yau manifolds the low energy effective Lagrangian
description breaks down and one could have additional light
fields.  The particular case of conifold was studied in \post .
In that case, say we have a vanishing $A$ cycle, and we turn
on a flux dual to that cycle.  This means we have a superpotential
$$W=\alpha a$$
where $a=\int _A\Omega$ and $\alpha= n_1+n_2\tau$. Let us set $n_2=0$.
If $A$ is shrinking we have in addition
a light wrapped D3 brane which in $N=1$ language corresponds
to chiral fields $\phi$ and $\tilde \phi$ with charge
$\pm 1$ respectively under the $U(1)$ gauge field
whose supersymmetric scalar partner has vev $\langle \phi
\rangle =a$. So the actual Lagrangian superpotential should be modified
to
$$W=\alpha a+a\phi \tilde \phi$$
Now the condition that $W=dW =0$ has a solution and is given by
$$a=0 \qquad \phi \tilde \phi =-\alpha$$
In fact it is possible to check that this actually preserves
the full $N=2$ supersymmetry. 
Even though some other singularities of Calabi-Yau
manifolds have also been considered in
\jm\ none
has been shown to lead to $N=1$ unbroken supersymmetry (though
a full no-go theorem is not available in this context).

\subsec{How to obtain $N=1$ supersymmetry?}
It thus seems difficult to obtain an $N=1$ supergravity solution
with $H$-flux turned on for compact Calabi-Yau manifolds.  
How could we possibly relax some conditions to make this possible?
The hint comes from considering $N=1$ supersymmetric Yang-Mills
theories.  In these cases one expects to have a mass gap with
some number of vacua $c_2(G)$ given by the dual Coxeter number of the group.
Moreover one can assign a superpotential to each vacuum given by
$$W_k=\omega_k exp (-S/c_2(G))$$
where $S=1/g^2$ and $\omega_k$ is an $c_2(G)$ root of unity.  
The meaning of this superpotential is that the domain
walls stretched going from one vacuum to the other will have a central
terms for their tension given by the difference of the corresponding
values of the superpotential.  In the usual $N=1$ Yang-Mills
the coupling constant $S$ is not a field but a parameter.  But
we can actually promote it to a chiral field whose vev is given
by the coupling constant.  If we do that, then we will
also have to consider $dW_k/dS =0$ for finding supersymmetric ground
states, otherwise we would get a positive energy given by
$$g^{S \bar{S}}|\partial_S W_k|^2$$
where $g^{S\bar{S}}$ is the inverse to the K\"ahler metric for $S$.  However,
$dW_k/dS =0$ has no solutions, which means that we have no supersymmetric
vacuum (or any vacuum in this case).  But clearly we can
embed the usual $N=1$ gauge theory in this theory, simply
by taking the kinetic term for $S$ field to be very large
and thus effectively freezing it (this corresponds to making
$g^{S\bar{S}}$ vanishing and thus giving no energy to the vacuum).
Note in this case that even if $S$ is treated as a parameter
the vacuum has an energy and so the supersymmetric background
makes sense if we decouple the gravity, by taking
 $M_{Pl}\rightarrow \infty $.
This is in fact how we will generate $N=1$ QFT's by
turning on $H$-fluxes; namely, as we will see later, the field
$S$ will play the role of an extra field, whose dynamics we will
freeze in the limit of interest and concentrate
on a decoupling limit where gravity is irrelevant.  This will in particular
avoid the no-go theorem of \jm\ for obtaining $N=1$ solutions .

\newsec{Geometric Engineering for $N=2$
 Theories and Their Partial Breaking to $N=1$}
In preparation of our discussion for turning on
RR-fluxes and breaking $N=2$ theories to $N=1$
 we first review some relevant aspects of geometric
engineering for $N=2$ gauge theories, in the context
of type IIA compactification on Calabi-Yau threefold
and its type IIB mirror\nref\kklmv{S. Kachru,
A. Klemm, W. Lerche, P. Mayr and C. Vafa, Nucl.
 Phys. B~459 (1996) 537.}\nref\klmvw{A. Klemm, W. Lerche, P. Mayr,
C. Vafa and N. Warner, Nucl. Phys. B~477 (1996) 746.}\nref\kkv{S. Katz,
A. Klemm and C. Vafa, Nucl. Phys. B~497 (1997) 173.}\nref\kmv{ S. Katz, P. Mayr
and C. Vafa, Nucl. Phys. B~497 (1997) 173.} (see \refs{\kklmv-\kmv}\
for more detail).  Instead of being
general, consider a simple example: Let us review
how the $SU(2)$ Yang-Mills is geometrically engineered:
We consider in type IIA a local CY geometry where
a $P^1$ is fibered over another $P^1$.  The simplest
possibility is $P^1_f\times P^1_b$.  In the limit
the $P^1_f$ goes to zero we obtain an enhanced
$SU(2)$ gauge symmetry from the $A_1$ singularity.
To suppress the gravity effects we take $g_s\rightarrow 0$.
The area of $P^1_b$, $S$, determines the coupling constant
for the $SU(2)$ gauge theory:
$$S={1\over g^2}$$
  We can identify
the size of $P^1_f$, $a$, with the vev of a scalar
in the adjoint of $SU(2)$; more precisely, we have
the classical relation that 
$$a^2=\langle Tr \phi^2 \rangle .$$
One considers the regime where $S$ is large and $a$
is small.  This is the same as taking a finite
size $P^1_b$ and zero size $P^1_f$ in the string frame in
 the limit
where $g_s\rightarrow 0$ (the effective mass of $D2$ branes
is given by $a=a'/g_s$ and $S=S'/g_s$ where 
the $a'$ and $S'$ denote the volume of the $P^1$'s in the
string frame).
In this limit the field $S$ becomes non-dynamical
and plays only the role of a parameter in the field theory.
The $N=2$ prepotential in this case has the following
structure
\eqn\preps{{\cal F}={1\over 2}Sa^2+a^2{\rm log}a^2+\sum_n c_n a^{2-4n}
{\rm exp}(-nS)}
for some constants $c_n$.  Note that $S$ can be absorbed
into a redefinition of $a$.  In fact this is the limit
one is taking in the geometry, namely we take $a\rightarrow 0$
and $S\rightarrow \infty$ keeping the combination
$a^4 {\rm exp}{S}$ fixed.
The way this expression is obtained is to use mirror symmetry
to compute the worldsheet instanton corrections in this
type IIA background by relating it to complex structure variation
in a type IIB background.  Physically, the net result is
that the apparent instanton corrections in \preps\
can be related to one loop corrections to the prepotential
summed over all D2 branes wrapped around the $P^1\times
P^1$ geometry.  In other words
$${\cal F}=Sa^2+\sum_{BPS\ m}m^2{\rm log}m^2$$
and the rich structure of the instanton corrections
gets mapped to an intricate structure of wrapped BPS
D2 branes in this geometry, in the limit we are taking.

\subsec{Adding the Flux and breaking $N=2\rightarrow N=1$}
Now we are ready to add the flux.  We choose the
flux to be an RR 2-form flux in the direction dual to the 2-cycle $P^1_b$.
This means according to our discussion in section 3, generating a
superpotential given by
$$W={\partial {\cal F}\over \partial S}$$
In the context of type IIB it means turning on a specific RR H-flux
dual to the 3-cycle representing $S$.  In that context $\partial
F/\partial S$ is a classical computation of periods. Now,
it has been shown \ref\mato{M. Matone, Phys. Lett. B~357 (1995) 342.}\ that
$${\partial {\cal F}\over\partial S} ={\rm constant}\cdot u$$
 where
$u=\langle tr \Phi^2 \rangle$, and 
the classical result $u=a^2$ receives quantum correction
exactly as given by $u={\rm constant}\cdot{\partial {\cal F} \over \partial S}$.
In particular, not just $a$, but also $u$ itself
is among the periods of the type IIB Calabi-Yau geometry (this
was a crucial fact for extracting gauge
theory implication of type IIB geometry \kklmv ).
We thus have a superpotential
$$W={\rm constant}\cdot u$$
This can be viewed in the gauge theory language
as a mass deformation (giving mass to the scalar
partner $\Phi$ of the vector multiplet) breaking $N=2 \rightarrow N=1$.
To find whether there are any supersymmetric vacua
one will have to analyze it, exactly as was done originally
in the field theory context by Seiberg and Witten \ref\seiwit{N. Seiberg and E.
Witten, \np 431 (1994) 484.}.
Namely one finds near the points where there is a massless dyon,
a modification of the superpotential, by including the light states.
In this case we get (somewhat analogous to the conifold case)
$$W= mu+ (a_D-a_0) \phi \tilde \phi$$
and one finds an $N=1$ supersymmetric solution 
$$\phi \tilde \phi =m {\partial u\over \partial a_D}\qquad a_D=a_0$$
In the type IIB context this corresponds to freezing
some complex moduli of Calabi-Yau threefold and rendering
some other moduli non-dynamical.

\subsec{Generalizations}
It is clear that this generalizes to a large number of
$N=2$ gauge models engineered in \kmv\ (at least
for the cases where the beta function is not zero).
  Just as was done
above for the $SU(2)$ case, one turns on in the type
IIA context independent RR 2-form flux in the bases whose
sizes control the coupling constants of various gauge groups.
By going over to its type IIB mirror, the RR flux turns
into a particular $H^R$ flux, which again serve to freeze
the moduli in the type IIB side, exactly as was done
for the $SU(2)$ case above.

\bigskip\noindent
{\bf Acknowledgements}~ We have benefited from discussions
with S. Ferrara, K. Hori, P. Mayr,
M. Porrati and A. Strominger. The work of T.R.T. was supported in 
part by NSF grant PHY-99-01057 and that of
C.V. was supported in part by NSF grant PHY-98-02709.

\listrefs
\end